\documentclass[review]{elsarticle}

\usepackage{lineno,hyperref,xcolor}
\modulolinenumbers[5]

\journal{Nucl. Instrum. and Meth. in Physics Research A,}

\begin{document}

\begin{frontmatter}

\title{Performance tests of boron-coated straw detectors with thermal and cold neutron beams}

\author{Georg Ehlers}
\address{Neutron Technologies Division, Oak Ridge National Laboratory, 
Oak Ridge, TN 37831-6466, USA}
\fntext[myfootnote]{This manuscript has been authored by UT-Battelle, LLC under Contract No. DE-AC05-00OR22725 with the U.S. Department of Energy. The United States Government retains and the publisher, by accepting the article for publication, acknowledges that the United States Government retains a non-exclusive, paid-up, irrevocable, world-wide license to publish or reproduce the published form of this manuscript, or allow others to do so, for United States Government purposes. The Department of Energy will provide public access to these results of federally sponsored research in accordance with the DOE Public Access Plan (http://energy.gov/downloads/doe-public-access-plan).}

\author{Athanasios Athanasiades, Liang Sun, Christopher S. Martin, Murari Regmi, 
Jeffrey L. Lacy}
\address{Proportional Technologies, Inc., 12233 Robin Blvd., 
Houston, TX 77045, USA}

\begin{abstract}
Prototypes of newly developed boron-coated straw (BCS) detectors have been tested in the thermal and cold neutron energy ranges. Their neutron detection performance has been benchmarked against the industry standard (detector tubes filled with $^3$He gas). 
The tests show that the BCS straws perform near their theoretical limit regarding the detection efficiency, which is adequate for scientific instruments in the cold neutron energy range. 
The BCS detectors perform on par with $^3$He tubes in terms of signal to noise and timing resolution, and superior regarding longitudinal spatial resolution. 
\end{abstract}


\end{frontmatter}


\section{Introduction}
\label{intro}

Neutron scattering is a powerful experimental technique with a wide range of applications in hard and soft condensed matter physics, chemistry, biology, and materials science~\cite{Mason06}.  
A number of international user facilities provide neutron beams for visiting researchers from government laboratories, academia, and industry, who perform thousands of scattering experiments every year~\cite{ILLRep2018,NISTRep2018}.

A critical component of every neutron scattering instrument is the neutron detection system.
Neutron detectors based on pressurized $^3$He gas tubes are in common use: 
the grand total of $^3$He gas in present installations at various neutron sources world-wide exceeds 90,000 liters.
The current global demand for $^3$He gas to use in {\em new} or {\em upgraded} neutron scattering instruments by far exceeds the available amount on the 
market~\cite{Zei12}.

The limited availability and high cost of $^3$He gas have motivated various development efforts for alternate detection technologies, with priority given to detection efficiency, spatial resolution, rate limit, cost, or any combination thereof~\cite{Tso11,Bell11,Big12,Rho12,Fuji12,Stef13,Naka14,Riedel15,Ana17,Pinto17,Wang18}. 
The needs of various types of neutron scattering instruments also differ in these performance metrics. 
For example, while single crystal diffraction instruments need to take a locally high rate and should possess very good spatial resolution, spectroscopy instruments emphasize efficiency, low intrinsic background and low sensitivity to $\gamma$-radiation.

To address the worldwide shortage of $^3$He, Proportional Technologies, Inc. (PTI) has developed the boron coated straw (BCS) neutron detection technology~\cite{Lacy11}. 
A $\sim{1}$~$\mu$m thick $^{10}$B-enriched boron carbide ($^{10}$B$_4$C) layer is uniformly sputtered onto a Cu foil, which is then cut to size and spiral welded along the length to form a straw with $\sim{7.5}$~mm diameter with the coating on the inside. 
BCS detectors are packed in groups of 7 inside sealed aluminum tubes. 
The tubes are 25.4~mm (1 inch) in diameter, with an active length of about 1~m, and allow sealed operation for indefinite periods of time. 
More details are given in section~\ref{details} below. 

When compared with other replacements for $^3$He gas, the BCS technology provides benefits such as a direct drop-in replacement for $^3$He tubes, stability, a neutron detection efficiency near the theoretical limit, a high tolerance to temperature extremes and shock, and very good $\gamma-$event rejection, resulting in an excellent signal to noise ratio. 
The readout electronics features a noise rejection algorithm which is essential to achieve proper resolution in the longitudinal direction of a tube.

\section{Design and Fabrication Details}
\label{details}

A single BCS detector contains a close-packed array of 7 thin walled, boron-coated copper tubes which are referred to as ``straws''. 
The straws are 7.5~mm in diameter, coated on the inside with a thin layer of $^{10}$B-enriched boron carbide $^{10}$B$_{4}$C. 
The $^{10}$B$_{4}$C layer is sputter coated onto a thin copper foil, and reels of this coated foil are spiral welded into their straw shape via an automated process. 
This BCS technology relies on the $^{10}$B$(n,\alpha)$ reaction: 
\begin{equation}
^{10}{\rm{B}} + n \rightarrow ^{7}{\rm{Li}} + \alpha{\;},
\end{equation}
wherein $^{7}$Li and $\alpha$ particles are emitted in opposite directions from the neutron absorption site with kinetic energies of 1.47~MeV and 0.84~MeV, respectively. 
For example, a $^{10}$B$_{4}$C layer of 1~$\mu$m thickness allows for one of the two charged particles to escape the layer 78\% of the time, ionizing the gas inside the straw. 
Therefore, the detection efficiency of a dense array of straws approaches 78\% at any desired wavelength as the number of layers increases.

\begin{figure}[t]
\includegraphics[width=4.5in]{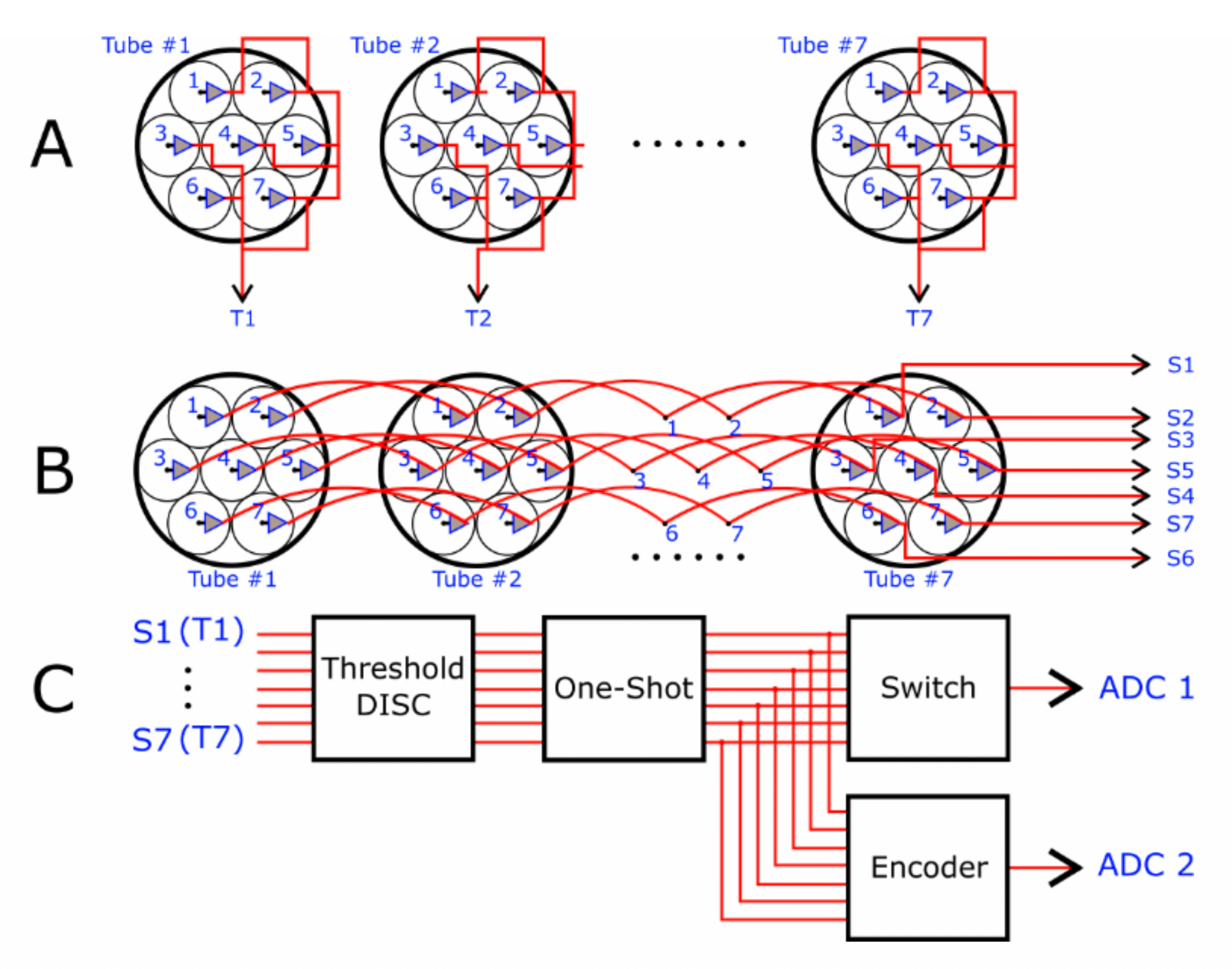}
\caption{\label{elec} Electronic readout schematics for an array of 7 tubes. A) One end of the detector panel outputs one signal per tube, designating each of the seven tubes as T1 through T7. B) The seven straws inside each tube are assigned numbers. The second end of the detector panel outputs one signal per straw designation as S1 through S7. In this way, the identity of the tube ($T_x$) and the straw within that tube ($S_y$) can be known. C) The seven outputs from either end of the 7-tube panel are passed through threshold discriminators prior to the shaping circuitry, followed by non-retriggerable one-shot multivibrators so that only one T and one S signal are digitized and read out as ADC signal per event. }  
\end{figure}

The schematics in Fig.~\ref{elec} shows how the 7 straws are close-packed in a single, sealed aluminum tube. 
The external tube is 25.4~mm in diameter, and can be up to 3.5~m long.  
Each tube is sealed using aluminum/ceramic end-caps, which allow for operation of the straw detectors in sealed mode. 
The detectors remain operational for indefinite periods of time in vacuum environments. 
One end-cap supports a gas port that allows for purging with the desired gas mixture to a pressure below 1~atm, after which the port is sealed.

The electronic readout of the detectors has been described in more detail elsewhere~\cite{Lacy2006,Lacy2011}. 
Briefly, the 7-straw tubes are decoded by connecting the 7 front-end amplifier outputs together within each tube on one end to identify the tube index, $T_x$ (Fig.~\ref{elec}A), and by connecting each straw to each same-numbered neighbor on the other end to identify the straw index, $S_y$ (Fig.~\ref{elec}B). 
For a given neutron event, only one signal from each is larger than a pre-set threshold value, providing the identification of the specific firing straw. 
The longitudinal position of neutron interaction is measured using a charge division method. 

The full readout scheme is modified from this simple description so that only one of the seven pre-amplifier outputs (at either end) containing the true neutron event is digitized, significantly reducing the readout hardware and increasing the count rate capability. 
This scheme is implemented by placing threshold discriminators (comparators) prior to the shaping circuitry, followed by non-retriggerable monostable multivibrators (one-shot) along with very fast very low resistance switch ICs, which open the appropriate gate to send only one of the seven with a signal to the Analog-Digital-Converter (ADC). 
A priority encoder is also implemented to transmit straw/tube ID to a buffer amplifier at either end which converts the code to one of seven pulse levels, subsequently digitized by another ADC. 
Since more than one switch can be turned on at once, control logic is designed to ensure that only one signal over threshold is selected. 
In addition, there is a latch to ensure that an over threshold signal arriving from other straws during the integration gate is prevented from causing the selector switches from changing. 
The readout block diagram is shown in Fig.~\ref{elec}C. 
Only four ADCs are required to read out a 7-tube group, a significant reduction in hardware cost.

\section{Experimental Setup For Detector Testing}

The Cold Neutron Chopper Spectrometer~\cite{CNCS11,CNCS16} (CNCS) at the Spallation Neutron Source (SNS) in Oak Ridge provided the environment for most of the detector tests. 
CNCS delivers a monochromatic pulsed beam of cold or thermal neutrons to the sample.  
The neutron energy (wavelength) range covered in the tests was from 0.5~meV (13~{\AA}) to 50~meV (1.3~{\AA}).  
The source frequency at SNS is 60~Hz and the pulse length on the sample is variable, in the range 
$\sim{20}-50$~$\mu$s.  
The pulse length on sample and the wavelength can be freely adjusted with the instrumental chopper settings.  
CNCS features a large array of 2~m long position-sensitive $^3$He tubes which are arranged in an arc at $\sim{3.5}$~m distance from the sample, covering about $180^{\circ}$ in the scattering plane from $\sim{+135}^{\circ}$ to $\sim{-50}^{\circ}$.  
The tube diameter is 1~inch and the partial $^3$He pressure is 6 bar.  
Tubes are vertically centered in the scattering plane and arranged in panels holding 8 tubes each (``8-packs'').  
The tube walls are 0.5~mm stainless steel (SS-304).  
The separation between the tubes is $\sim{1}$~mm to allow the placement of neutron absorbing fins between the tubes.  
In order to avoid parasitic air scattering, the detector array is located in a large enclosure that is filled with Ar gas at ambient pressure mixed 
with $\sim{2.25}$\% of CO$_2$.  

The tests lasted several months and addressed the following: detection efficiency of the BCS prototype detector, timing resolution, signal peak to background ratio, and long term stability.  
The tests were conducted in parallel with normal instrument operation (primarily SNS user experiments).  

\begin{figure}[t]
\includegraphics[width=4.5in]{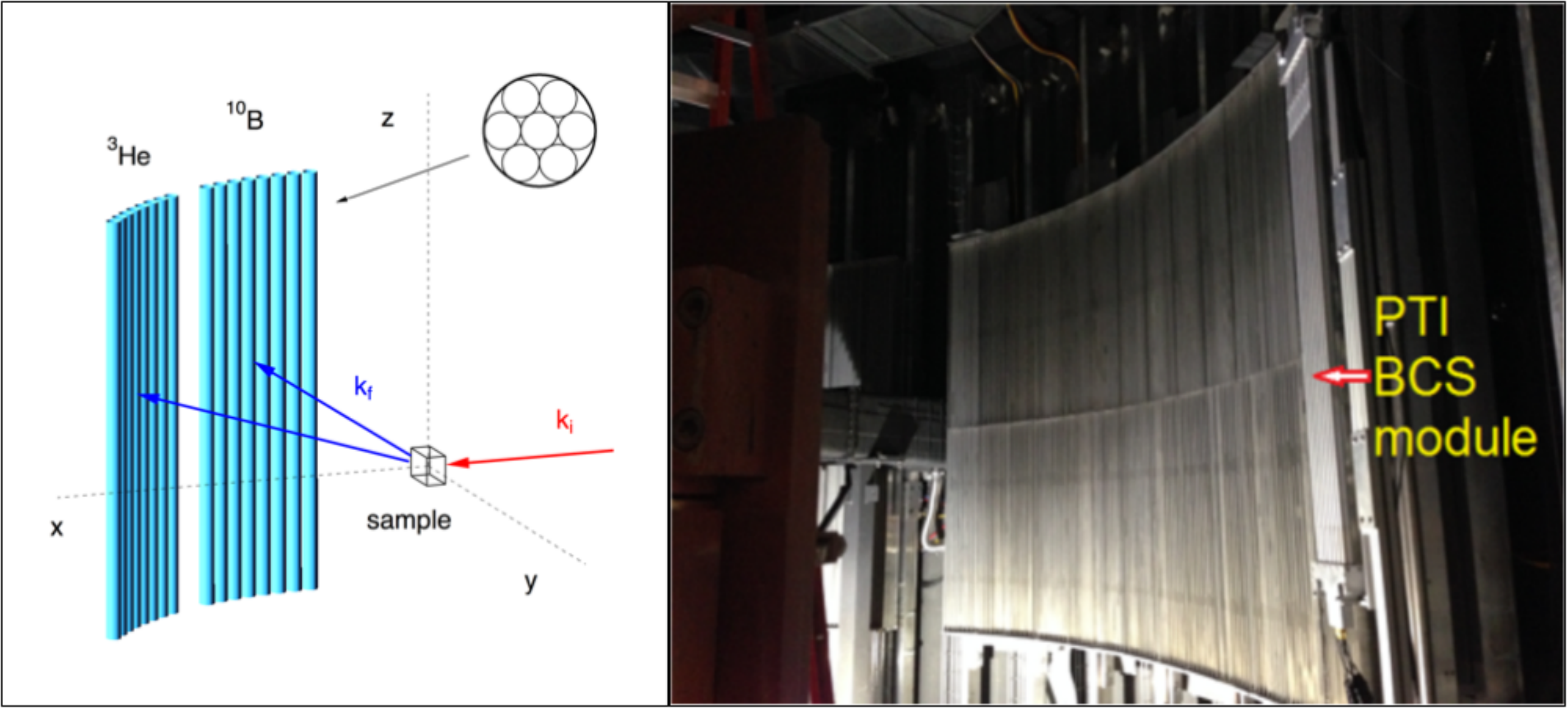}
\caption{\label{setup} Left: schematic setup of the prototype testing at CNCS. Right: photograph of the installation with the 1-layer prototype. The BCS detector is to the right of the $^3$He tube array. }  
\end{figure}

Two BCS prototype detectors were tested at CNCS. 
The first was a single-layer panel consisting of seven 1~m long aluminum tubes with a diameter of 1~inch. 
This setup is shown in Fig.~\ref{setup}.  
Each BCS tube holds 7 straws as shown in the top center inset of Fig.~\ref{setup}. 
The 1-layer panel was centered vertically in the same way as the $^3$He tubes. 
The panel was mounted directly at the side of the last 8-pack of $^3$He tubes at a scattering angle of $\sim{-53}^{\circ}$. 
The BCS tubes were located slightly closer to the sample (3.194~m) than the $^3$He tubes (3.478~m) of the nearest 8-pack, which was the one primarily used for performance comparisons. 
To this end, the absorbing fins between the $^3$He tubes were removed from 
this 8-pack. 
The back and sides of the 1-layer panel were shielded with neutron absorbing Cd. 
The CNCS $^3$He tubes also have such shields on the backend.

From October 2013 until May 2014, the 1-layer panel was operated for more than 2500 hours logging $>{200}$ million events.  
Straw performance was totally reliable over the entire testing period with no observed degradation of performance in any of the 49 BCS detectors.  
The only problem encountered was leakage experienced in the external HV electrical boards at highest voltages due to the Argon/CO$_2$ atmosphere employed in the instrument. 
This was remedied by appropriate conformal coating in the design of the second prototype.

\begin{figure}[t]
\includegraphics[width=4.5in]{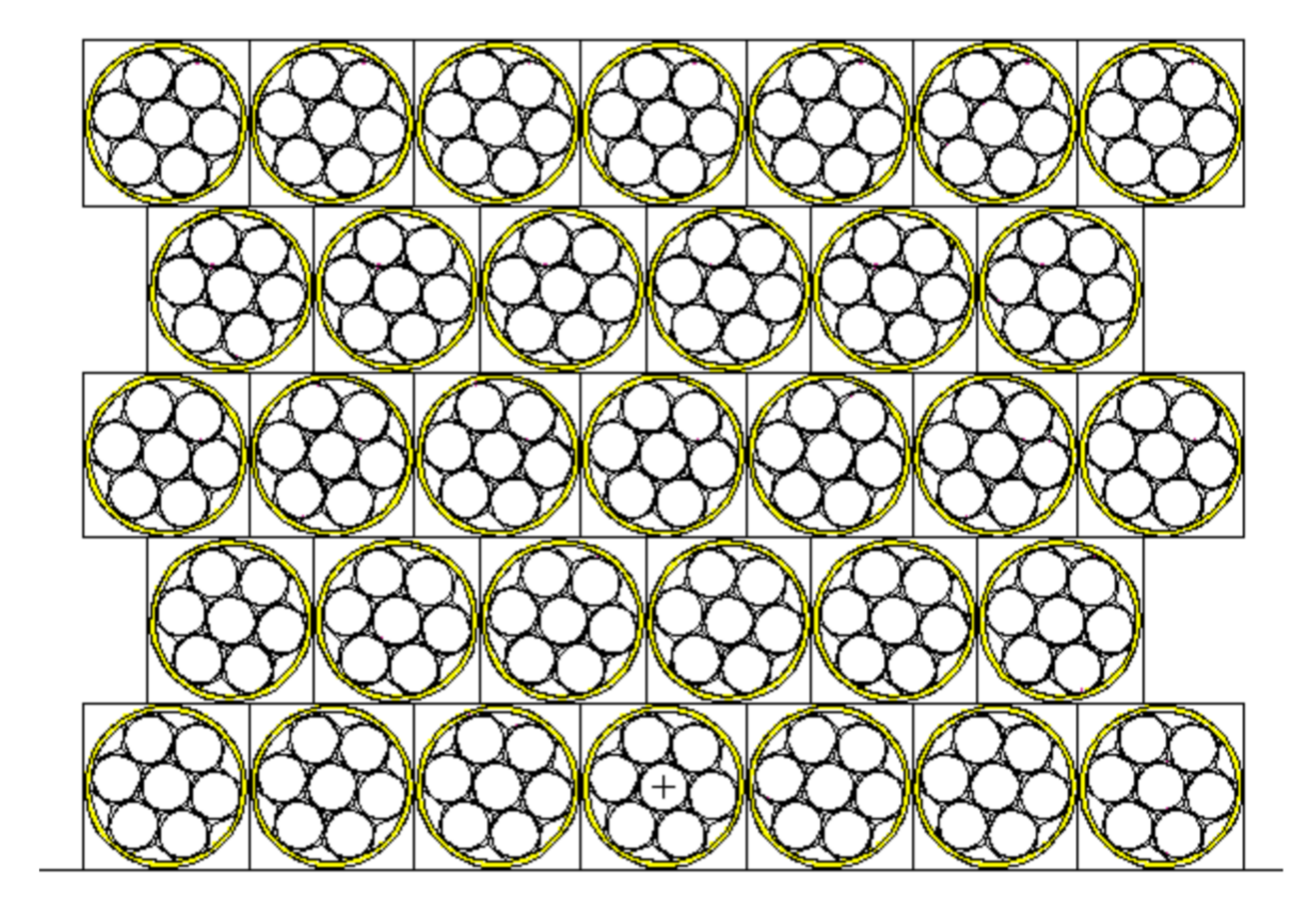}
\caption{\label{multi} Grouping of BCS detectors in the second 5-layer prototype  panel. Each 1-inch aluminum tube contains 7 straws. }  
\end{figure}

The second prototype detector was a multi layer panel which was an arrangement of 33 tubes in 5 layers in a 7-6-7-6-7 arrangement (see Fig.~\ref{multi}). 
The main purpose of this second generation prototype was to achieve higher counting efficiency. 
This panel was tested from January to May 2016. 
As before, the panel was mounted right next to the last 8-pack of $^3$He tubes.
The distances of the individual layers to the sample position were calibrated by neutron time-of-flight, see below. 
Similar neutron background shielding with Cd was used for this panel.

The 1-layer panel was also tested at the CG-1A beam line at the HFIR reactor at Oak Ridge. These tests focused primarily on the spatial resolution (longitudinally) of the individual tubes. 
A monochromatic beam with a wavelength of $\lambda=4.22$~{\AA} was used in these tests. 

\section{Results}

\subsection{General comparison}
\begin{figure}[t]
\includegraphics[width=4.5in]{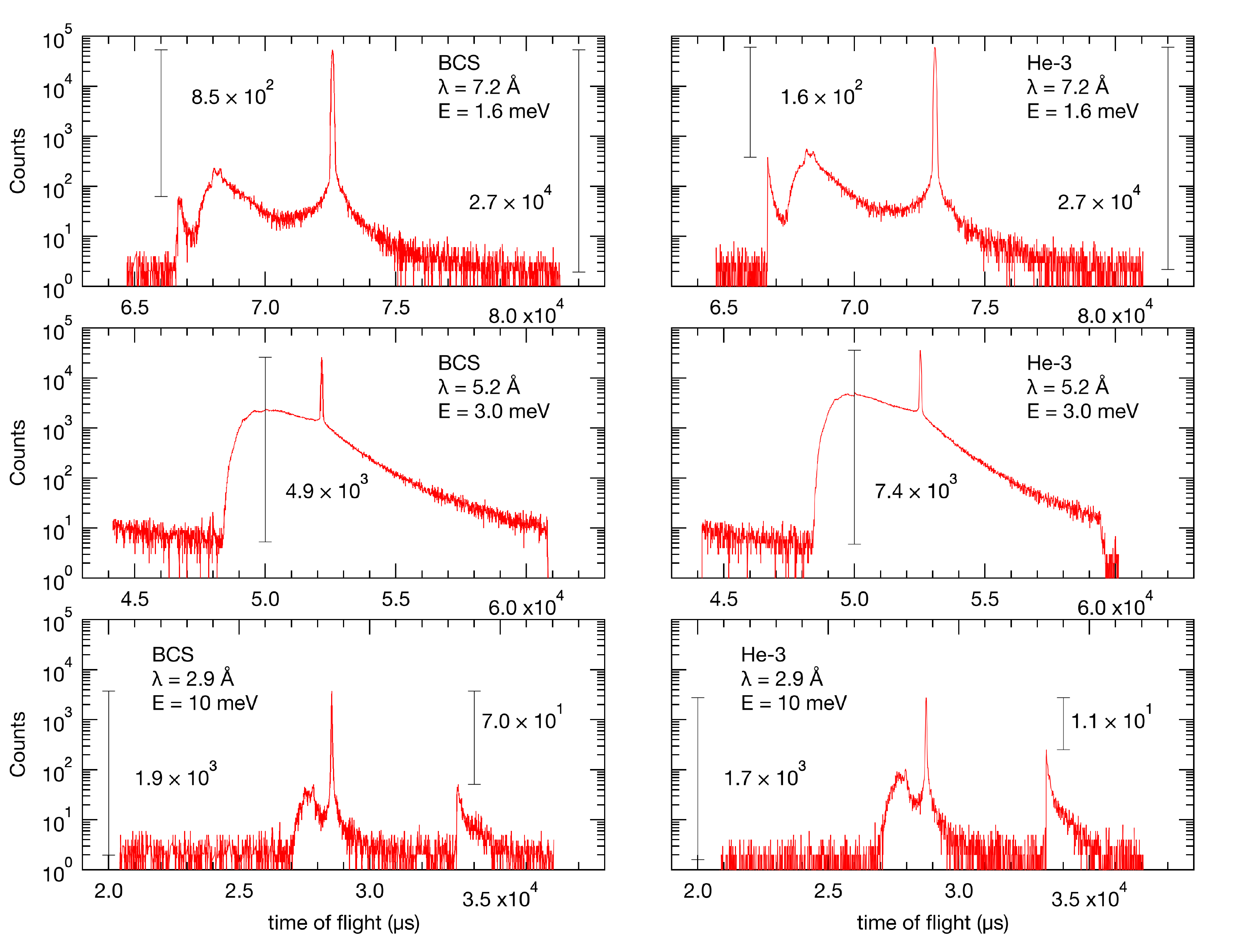}
\caption{\label{tof1} Time-of-flight spectra measured with the 1-layer BCS detector and the $^3$He tubes at three different energies. Counts are not normalized in the same way between the two detectors and therefore not directly comparable. 
The neutron energies are (top to bottom) 1.6~meV, 3.0~meV and 10~meV, respectively.  
The actual count rates in these measurements are $\sim{1-10}$ counts per second per meter of tube. 
The shapes of corresponding curves between the two detectors are not identical because the distance to the sample was not the same. This can also be seen in the position of the elastic line which is slightly shifted towards longer time with the helium tubes.}  
\end{figure}

Time-of-flight spectra measured with the 1-layer BCS detector and the $^3$He tubes are shown in Fig.~\ref{tof1}. 
These spectra show raw data binned in time-of-flight without any processing except time stamping. 
These measurements were made with different samples at different times. 
Therefore, the intensity scales in these plots cannot be directly compared.
These spectra show convincingly several key properties of the BCS detectors, regarding: 
\begin{enumerate}
\item the peak intensity relative to the apparent ambient background level
\item dynamic range
\item the elastic peak to flash pulse height ratio (see below)
\end{enumerate}
The performance of a detector in these metrics is of great importance for a time of flight instrument. 
The ratio of the peak intensity to the apparent ambient background level is nearly the same for both detectors. 
The numbers are directly comparable because both detectors were shielded in the same way with Cd on the backend (see above).
It has long been recognized that the low efficiency for $\gamma-$radiation is one key property of detectors based on $^3$He. 
Spectra measured at 1.6 meV and 10 meV feature the flash pulse of 
high energy ($\sim{1}$~MeV) neutrons which is emitted by the source when the proton pulse hits the target. 
These neutrons propagate through many meters of shielding and arrive at the detector at virtually zero true time of flight which is why they occur at integer multiples of 16,666.7~$\mu$s (after folding into the appropriate frame by the data acquisition software). 
The peak to flash pulse height ratio for the two detectors is nearly the same as well (the BCS detector actually performs better than the $^3$He tubes in this regard). 

The spectra at the three energies were measured with different samples at different times. The intensity at 1~meV and 10~meV was rather low in absolute terms which is why the flash pulse is more prominently visible. The spectra at 3~meV were measured with a protonated soft hydrogenated material and contain strong inelastic scattering all the way to infinite energy gain 
(which occurs at $\sim{48,000}$~$\mu$s in this case).

The scattering around the elastic time of flight (``quasi-elastic scattering'') is often of particular importance. 
The occurrence of such scattering reveals the existence of dynamic modes in a material that are very slow on average or have a low energy scale (much lower than the energy of the neutron). 
Diffusive motions fall into this category (for example, of particles suspended in a liquid), or the relaxational dynamics of molecules of a viscous liquid. 
The two most important characteristics of an instrument for such studies are the energy resolution (width of the elastic line) and the instrument background level around the elastic line. 
Quasi-elastic scattering will be seen as a broadening of the elastic line or as an increase in scattering around it, or both.
The sharper the elastic line and the lower the instrument background around it, the smaller the effects one can discern in a real material.

\begin{figure}[t]
\includegraphics[width=4.5in]{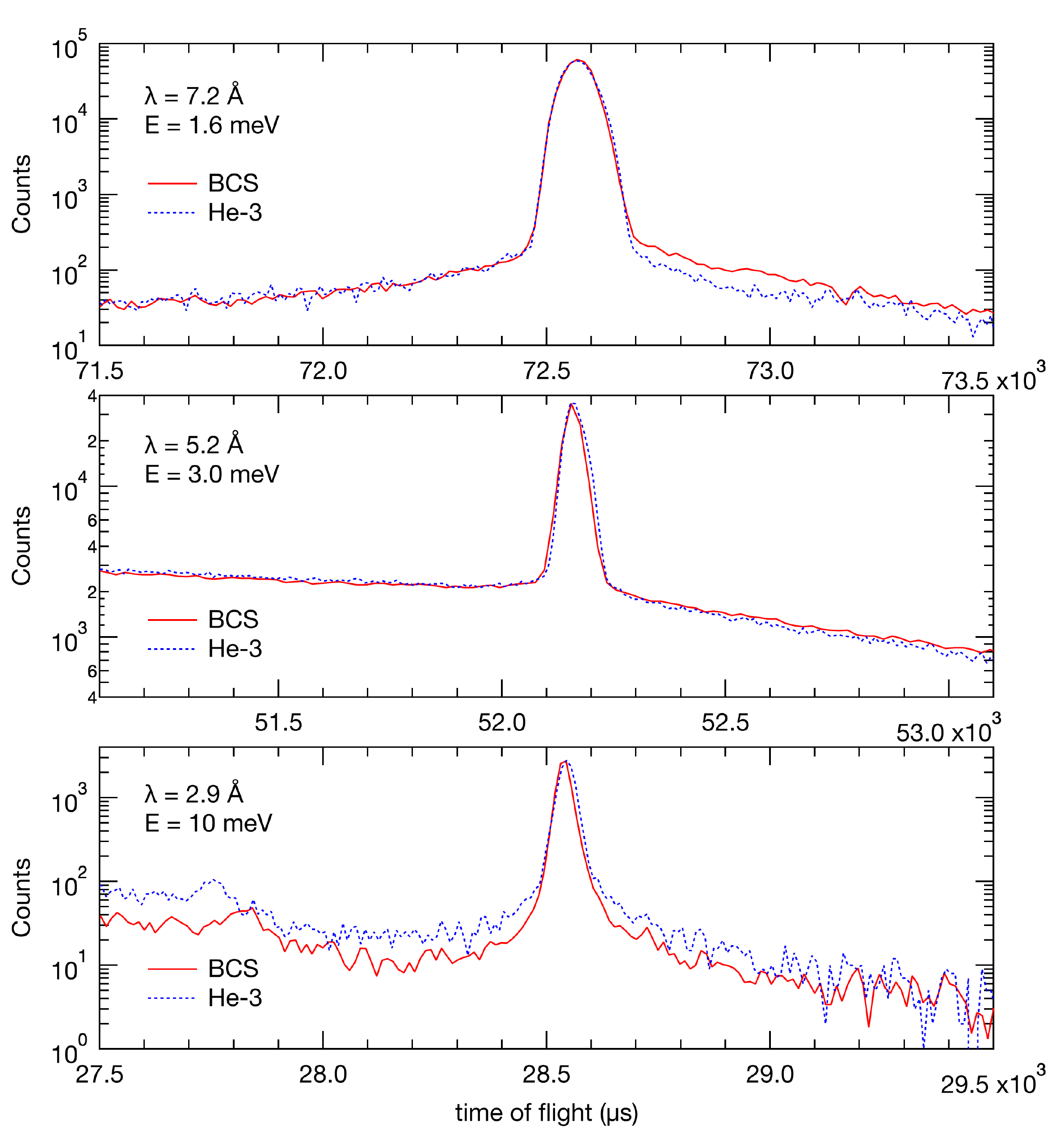}
\caption{\label{tof3} Time-of-flight spectra measured with the 1-layer BCS detector around the elastic time of flight. With not fully optimized detector shielding the BCS detector fared better than the $^3$He tubes at thermal energies (10~meV) but somewhat worse at low energy (1.6~meV).}  
\end{figure}

The datasets from Fig.~\ref{tof1} are again shown in Fig.~\ref{tof3}, zooming into the range around the elastic time of flight.
The 3~meV measurement is actually not particularly conclusive for the detector performance in this area because the sample was a strong scatterer in the quasi-elastic regime. 
However, both the 1.6~meV and the 10~meV measurements show that the BCS detector was on par with $^3$He tubes in this range. 
To help with the comparison, the BCS data were translated horizontally on the time axis to overlap with the $^3$He tube data.

\begin{figure}[t]
\includegraphics[width=5in]{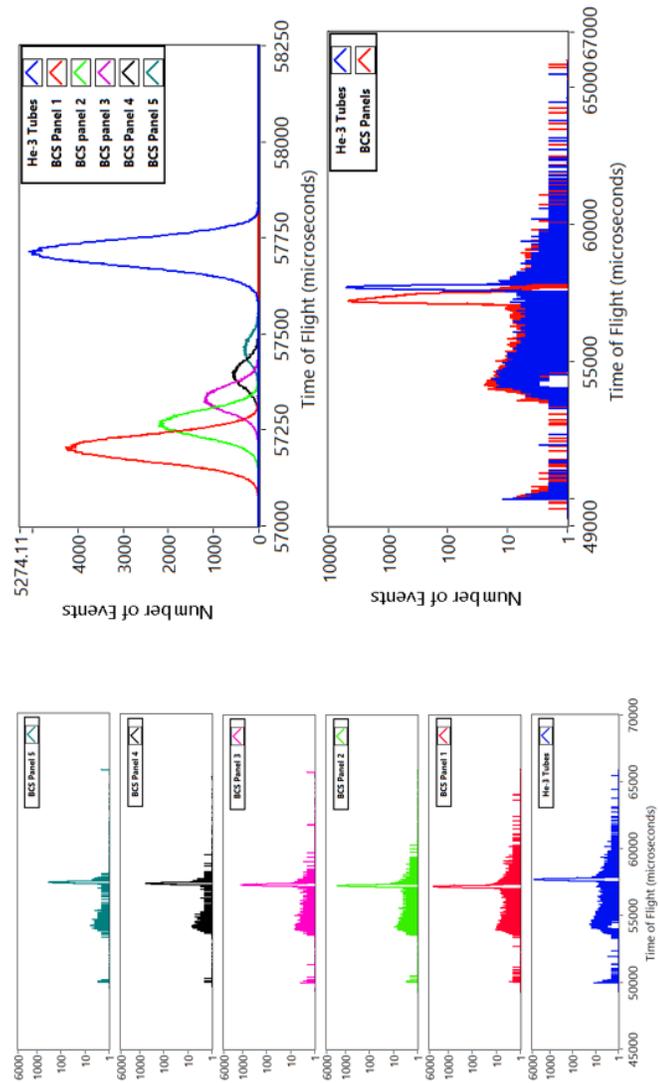}
\caption{\label{tof5} Time-of-flight spectra measured with the 5-layer BCS detector and the $^3$He tubes at 2.49~meV incident energy. The five layers of the BCS detector are read out individually. Because the data are plotted as of time of flight, the peak width is much larger for the BCS detector given its larger depth. However, the energy resolution of the two detectors is virtually the same.}  
\end{figure}

Time-of-flight spectra measured with the 5-layer BCS detector and the $^3$He tubes are shown in Fig.~\ref{tof5}. 
The five layers are separated out individually, and the position of the elastic time of flight peak indicates the different distances of the layers to the sample. 
One can also see how the layers count less and less indicating a shadowing effect, which is expected. 
It remains to be analyzed how much this matters for real neutron scattering applications. 
Another concern is the secondary scattering in the aluminum which matters at energies above the Bragg cut-off at 3.74~meV. 
Again, it will require more dedicated testing to investigate this issue. 
The pictures show that a depth of five active layers is about adequate for cold neutrons. 
Comparing time-of-flight spectra, in particular, the ratio of the count rate at the elastic line to the apparent time-independent background, reveal that the straw detector performs equally well (if not better) than the $^3$He tubes.

The next generation of straw detectors will address concerns of secondary scatter in detector materials by introducing design changes, to include straw walls made from high-purity aluminum, rather than copper, and containment tubes with a thinner wall of 0.25 mm. Initial production runs showed that boron could be successfully coated onto 25-$\mu$m-thick aluminum foil; the coated foil was then formed into a straw, using the same spiral welding technique, with better results compared to the copper foil. At the same time, containment tubes employed in current production were successfully machined down to an outer diameter of 24.1 mm, from 25.4 mm (1.00 inch), reducing the  wall thickness down to 0.25 mm (from 0.89 mm). The above design changes  reduce the mass of detector materials by a factor of 3.5, with a corresponding reduction in scatter.

\subsection{Counting Efficiency} 
\begin{figure}[t]
\includegraphics[width=4.5in]{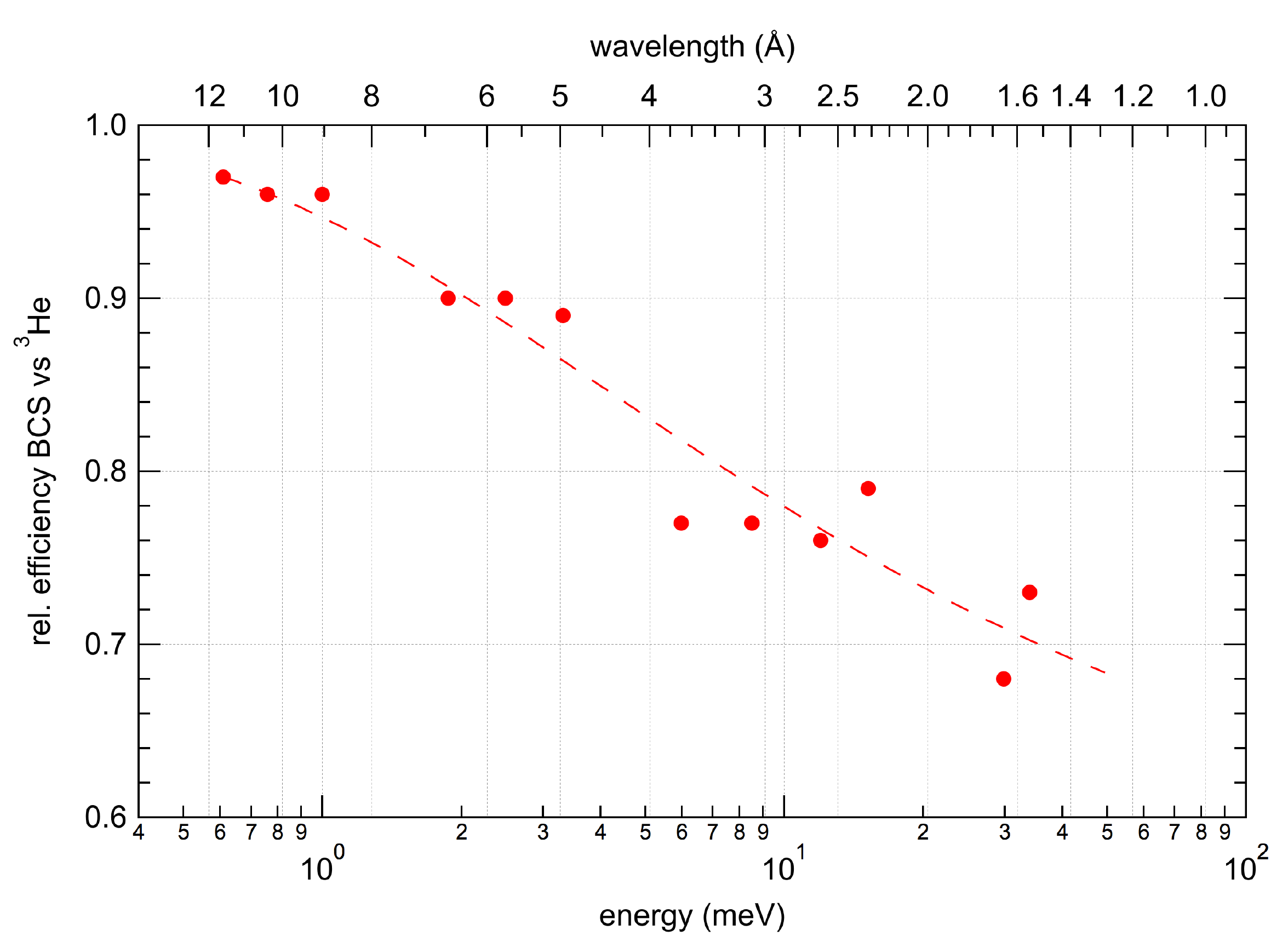}
\caption{\label{eff} Efficiency of the 5-layer BCS detector against the $^3$He tubes measured at CNCS. The dashed line is a guide to the eye. }  
\end{figure}

One of the most important metrics for neutron scattering applications is the efficiency with which neutrons are detected. 
The performance of the BCS 5-layer detector against the $^3$He tubes is shown in 
Fig.~\ref{eff}.  
A vanadium standard sample was used in these measurements and only the elastic scattering was used for the analysis. 
The count rates in each detector were measured simultaneously, and were corrected for the detection solid angle in order to achieve a fair comparison. 
The efficiency is within 10\% of expected values calculated by MCNP code.

\subsection{Longitudinal Spatial Resolution}
\begin{figure}[t]
\includegraphics[width=4.5in]{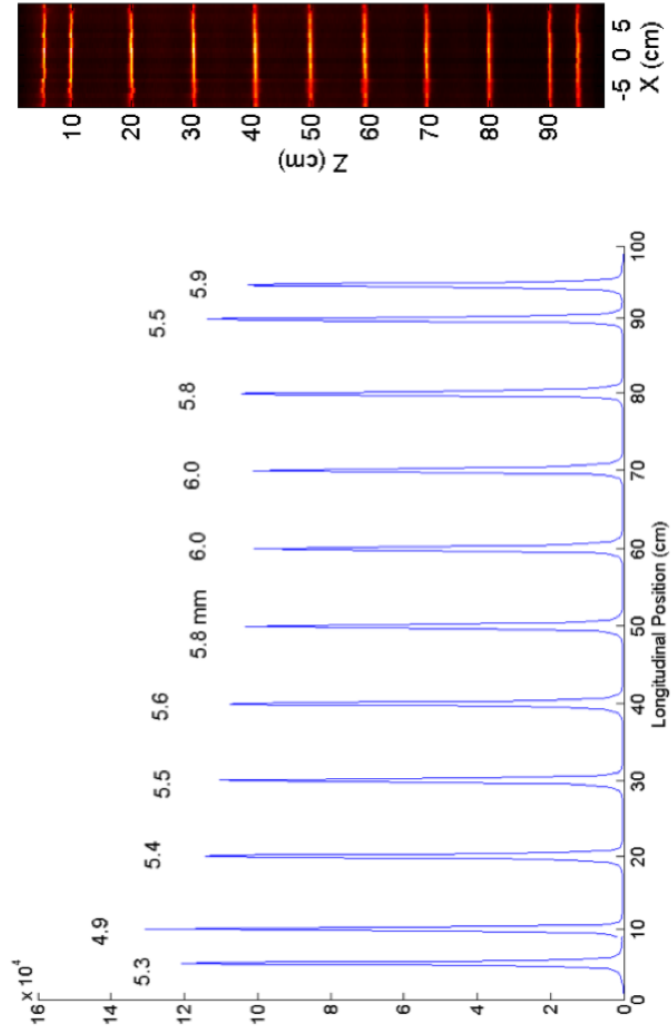}
\caption{\label{reso} Spatial resolution measurement with monochromatic beam at HFIR. Individual peak widths (full width at half maximum) are indicated in the figure.}  
\end{figure}

For these tests the 1-layer panel was translated vertically behind a 1~mm wide slit that was made of neutron absorbing material. 
The main result is shown in Fig.~\ref{reso}. 
Measurements were taken at 11 longitudinal locations (5~cm, 10~cm, 20~cm, 30~cm, 40~cm, 50~cm, 60~cm, 70~cm, 80~cm, 90~cm and 95~cm).  
Data were collected continuously during the full translation at each longitudinal location.  
Fig.~\ref{reso} shows longitudinal resolution at 11 locations as well as a reconstructed image using a superposition of all collected data.  
The resolution is $5.5\pm{0.4}$~mm full width at half maximum (FWHM) and is quite uniform along the length excluding small regions at the ends of the tubes of less than 5~cm.  
The noise rejection in the readout electronics is essential to achieve this resolution.

\section{Conclusion}
Prototypes of newly developed boron-coated straw detectors have been tested in a cold neutron beam. 
Their performance has been benchmarked against the industry standard of detector tubes filled with $^3$He gas. 
The tests show that the BCS detectors perform on par with $^3$He tubes in terms of signal to noise ratio and timing resolution. 
They are better than $^3$He tubes regarding spatial resolution both in the longitudinal direction and horizontally (because of the smaller diameter of the straws). 
Regarding the detection efficiency of cold neutrons, the BCS straws are near their theoretical limit, and perform adequately in the cold neutron energy range.

\section{Acknowledgments}

We would like to thank Rick Riedel for his support of the tests at HFIR, and Jana Olson for critical reading and comments. 
This work was supported by the U.S. Department of Energy (DOE), under SBIR Award No. DE-SC0009615.
A portion of this research used resources at the High Flux Isotope Reactor and the Spallation Neutron Source, which are DOE Office of Science User Facilities operated by the Oak Ridge National Laboratory.


\bibliography{References}

\pagebreak

\pagebreak

\pagebreak

\pagebreak

\pagebreak

\pagebreak

\pagebreak

\pagebreak

\end{document}